\documentclass[%
 reprint,
 amsmath,amssymb,
prstab,
]{revtex4-1}

\usepackage{graphicx}
\usepackage{dcolumn}
\usepackage{bm}

\begin{document}

\preprint{APS/123-QED}
\title{Knife-edge based measurement of the 4D transverse phase space of electron beams with  picometer-scale emittance}

\author{Fuhao Ji$^{1}$, Jorge Giner Navarro$^{2}$, Pietro Musumeci$^{2}$, Daniel Durham$^{3,4}$, Andrew Minor$^{3,4}$, and Daniele Filippetto$^{1, }$}
\email{dfilippetto@lbl.gov; }
\affiliation{$^{1}$ Accelerator Technology and Applied Physics Division, Lawrence Berkeley National Laboratory, One Cyclotron Road, Berkeley, California, 94720, USA \\
$^{2}$  Department of Physics and Astronomy,
University of California, Los Angeles,
Los Angeles, California 90095, USA \\ 
$^{3}$ Department of Materials Science and Engineering,
University of California, Berkeley, 
Berkeley, California 94720, USA \\
$^{4}$ National Center for Electron Microscopy, Molecular Foundry, Lawrence Berkeley National Laboratory, One Cyclotron Road, Berkeley, California, 94720, USA\\}

\date{\today}

\begin{abstract}

Precise manipulation of high brightness electron beams requires detailed knowledge of the particle phase space shape and evolution.
As ultrafast electron pulses become brighter, new operational regimes become accessible with emittance values in the picometer range, with enormous impact on potential scientific applications.
Here we present a new characterization method for such beams and demonstrate experimentally its ability to reconstruct the 4D transverse beam matrix of strongly correlated electron beams with sub-nanometer emittance and sub-micrometer spot size, produced with the HiRES beamline at LBNL. Our work extends the reach of ultrafast electron accelerator diagnostics into picometer-range emittance values, opening the way to complex nanometer-scale electron beam manipulation techniques.

\end{abstract}

\pacs{Valid PACS appear here}
\maketitle


\section{\label{sec1}Introduction}


The advent of ultrafast lasers and rapid development of particle accelerator technology paved the way to the generation of dense, ultrashort electron pulses. Using time-varying electromagnetic fields in the radiofrequency range, electron beams can be rapidly accelerated to relativistic energies and longitudinally compressed down to the single-digit femtosecond durations \cite{MusumeciShort}. Similarly, the peak beam transverse brightness greatly benefits from the smaller and denser volumes in the transverse trace space generated with the help of large field amplitudes at emission plane \cite{FilippettoBrigthness}\cite{BazarovBrigthness}. Moving one step further and coupling high fields with MHz repetition rates results in a leap in average electron flux \cite{APEX}\cite{HiRES}, which can in turn be used to produce transverse emittance values in the nanometer and picometer range \cite{BrightSource}\cite{Li2012}\cite{nanoUED}, with a potentially enormous impact on scientific applications, including free-electron lasers (FEL) \cite{Ackermann}\cite{Emma}, ultrafast electron diffraction (UED) \cite{Zewail}\cite{SLACUED} and microscopy (UEM)\cite{UEM}, injection into laser-driven microstructure accelerators \cite{England} \cite{Stragier}, inverse Compton scattering\cite{Graves}, and high average power THz generation \cite{Shen}. 
In order to take full advantage of the small transverse phase space generated, accurate knowledge of the four-dimensional (4D) transverse beam matrix is essential. 
A variety of different  diagnostics techniques have been developed to measure transverse beam properties, including quadrupole/solenoid scan techniques \cite{Prat}, pepper pot \cite{PepperPot} and slit scan\cite{slitscan} methods. 

Quadrupole/solenoid scans are widely used for emittance measurements. The beam dimensions at a fixed point along the beamline are recorded as function of the strength of upstream electron optics, retrieving the full 4D beam matrix \cite{Prat}. In typical setups the beam projection in the \textit{(x,y)} plane is measured via optical imaging methods providing a resolution of few micrometers at best. 


The pepper pot method\cite{PepperPot} uses an electron mask for sampling the beam transverse phase space at multiple positions simultaneously. It is therefore a natural choice for retrieving single-shot information. Indeed the shadowgraph of transmitted beamlets on a downstream screen carries information on the coupled four-dimensional beam matrix. 
The trade-off between signal-to-noise ratio (SNR), mask aperture size ($\geq10~\mu m$), distance between mask and detector ($\sim m$), and imaging system spatial resolution and efficiency determines the final resolution, practically limiting it to emittance values  in the few-nanometer range.
Recently the TEM grid method\cite{Li2012}\cite{Jorge} has been introduced as an alternative to pepper pot. By analyzing the the positions and sharpness of the grid bars, as well as the intensity of the image, the entire transverse beam matrix at the grid plane can be reconstructed. Unlike the pepper pot, here a large portion of the beam is transmitted trough the holes in the grid leading to higher contrast (SNR) and accuracy of the measurement. The shadowgraph of the TEM grid is used to retrieve the beam properties, requiring a beam waist before the grid and a beam size of at least few grid holes at the grid. Measurements of beams close to the waist or with strongly tilted ellipses in the phase space are subject to large errors \cite{Jorge}. 

In general the smallest emittance values measured so far with the techniques described above are of the order of few nanometers~\cite{Li2012}. 




\begin{figure}
\includegraphics[width=8cm]{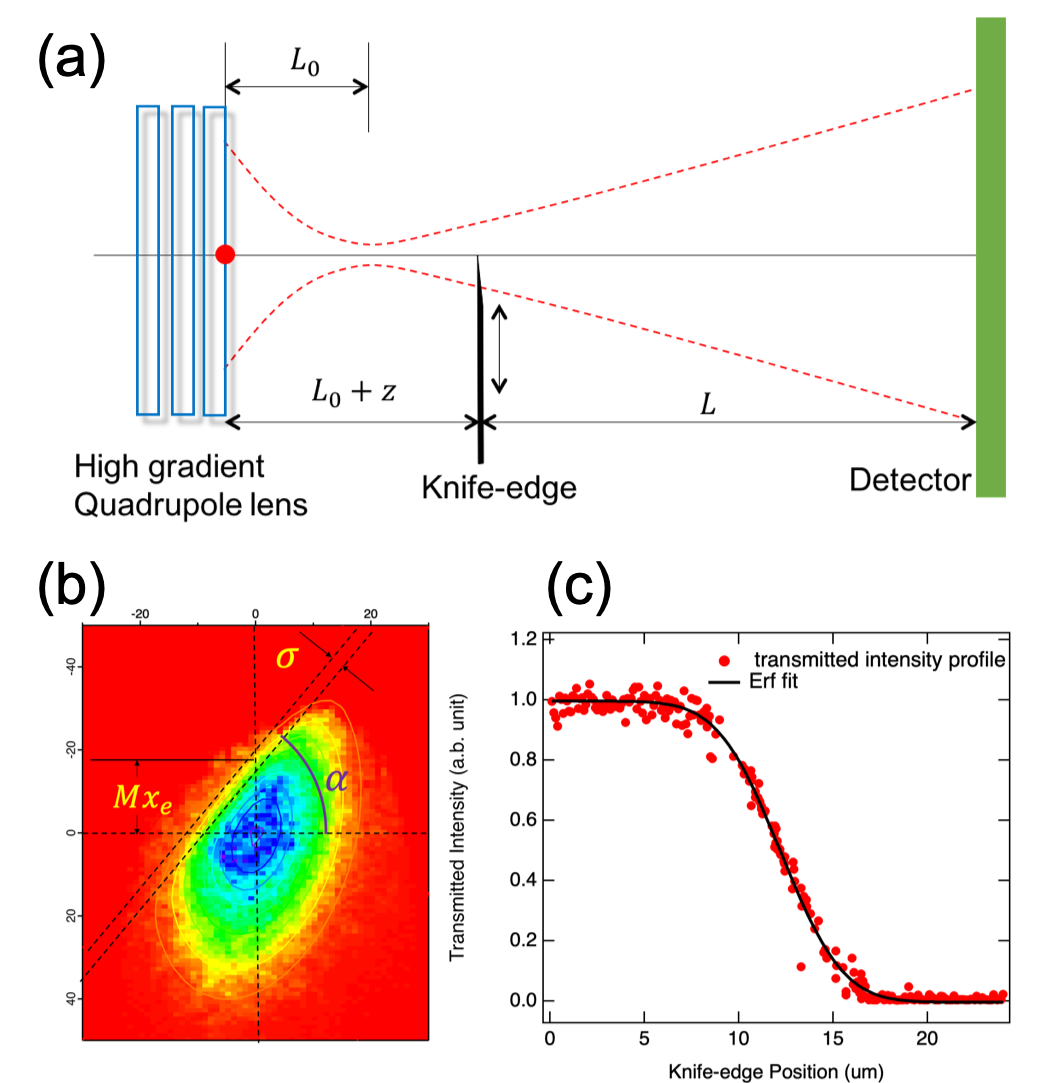}
\caption{\label{figure1} (a) Sketch of the knife-edge scan setup. Red dot indicates the reconstruction point, and red dashed lines represent beam edges. (b) A typical knifed beam image on the detector, note that the projection of the cut undergoes a rotation of angle $\alpha$ after propagation (c) The transmitted intensity profile in the knife-edge scan, Erf fitting gives the rms beam size at the knife-edge plane}
\end{figure}

In this work, we introduce a new methodology for characterization of the four-dimensional transverse beam matrix of electron beams extending the reach of the measurement space into picometer-scale  emittance values and nanometer-scale spot sizes.  The technique merges the methodology typical of quadrupole scan described above with the high spatial precision in beam size measurements given by the knife-edge scan technique (widely used in laser optics\cite{KE}) assisted by a powerful data analysis and global fitting routine. The knife-edge scan measurement technique is conceptually simple. By moving a knife-edge obstacle into the electron beam with nanometer-scale precision  on the edge positioning,  sub-micrometer spot sizes can be reliably measured by detecting the amount of electrons surviving the obstacle. The main idea behind our work is to go one step further and utilize all of the information in the actual image generated by the knife-edge instead of only measuring the transmitted charge.


We describe the concept, simulation results, experiments as well as the data analysis of an example measurement. This work was conducted at the High Repetition-rate Electron Scattering (HiRES)\cite{HiRES} beamline at LBNL. Our results show the flexibility and the potential for such technique as high accuracy tool for measuring the evolution of transverse 4D phase space beam matrix sub-micrometer beam size and picometer range emittance, extending the reach of ultrafast instrumentation techniques by more than one order of magnitude in the transverse space. 

The paper is organized as follows: In Section II, the knife-edge scan technique is described and formalized. In Section III, a numerical simulation of virtual measurement is presented. In Section IV, application of knife-edge scan technique at the HiRES beamline is described. The experimental results and detailed data analysis are presented. In Section V we summarize the work and discuss possible future applications of the technique in the R\&D of ultrahigh brightness electron sources

\section{\label{sec2}Theoretical model}

The experimental setup consists of an electron-opaque metal obstacle with a sharp edge ($\leq10~nm$ RMS roughness) intercepting the beam. The edge is moved through the beam using a calibrated translation stage with 100 nm resolution and the transmitted electrons are imaged 0.6~m downstream by a scintillator and an intensified CCD camera (see Fig.~\ref{figure1}). The resulting series of beam images carry information on the beam size at the obstacle position, the local angle distribution as function of edge position along the beam, and the beam correlations in the four-dimensional transverse phase space \textit{(x,x',y,y')}.


Assuming a Gaussian beam distribution, the total transmitted charge as function of edge position along the $x$ direction has the shape of a cumulative distribution function:
\begin{equation}
Q(x_e)=\frac{Q_0}{2}\left[1+\textrm{erf}{\left(\frac{x_e-x_0}{\sqrt{2} \sigma_{ke,x}} \right)}\right],
\label{eq:charge_scan}
\end{equation}
where $Q_0$ is the total charge, $x_e$ and $x_0$ are respectively the edge position and the beam center position, and $\sigma_{ke,x}$ is the beam size at the knife-edge plane. Figure~\ref{figure1}(c) shows an example of measurement, where the intensity change recorded as function of the knife-edge position is used in a least-square-fit procedure using Eq.\ref{eq:charge_scan}.  


In the general case of electron beams with correlated transverse planes, the knife-edge method presented allows the reconstruction of the entire 4D beam matrix $\Sigma_{4D}$:
\begin{equation}
	\Sigma_{4D}=
	\begin{pmatrix}
		\left<x^2\right> & \left<xx'\right> & \left<xy\right> & \left<xy'\right> \\
		\left<xx'\right> & \left<x'^2\right> & \left<x'y\right> & \left<x'y'\right> \\
		\left<xy\right> & \left<x'y\right> & \left<y^2\right> & \left<yy'\right> \\
		\left<xy'\right> & \left<x'y'\right> & \left<yy'\right> & \left<y'^2\right>
	\end{pmatrix},
\end{equation}
and the 4D emittance:
\begin{equation}
\epsilon_{4D} = \sqrt{\det{\left( \Sigma_{4D} \right)}}
\label{4Demittance}
\end{equation}

As described below in detail, in this case recording only the transmitted intensity is not enough, but rather the information of the entire projection on the \textit{(x,y)} plane is needed. 
The full reconstruction of the transverse beam matrix assumes a particularly important role in the case of electron beams tightly focused to sub-micrometer scales, as it allows to measure and diagnose non-idealities in the focusing systems, such as  astigmatism and other aberrations. For example, if a slightly astigmatic system is used on a beam with correlated transverse planes, then measuring separately the beam size along two orthogonal directions provides values always larger than the actual waist size.

Let us consider a Gaussian distribution in the 4D transverse phase space that is partly intercepted by the knife-edge at the position $x_e$, such that only the particles with $x>x_e$ are transmitted to the detector. After drifting for a distance $L$ downstream the obstacle, the distribution limit is defined as: 
\begin{widetext}
\begin{equation}
	\rho(x,x',y,y') = 
	\left\{
					\begin{matrix}
            \frac{Q_0}{4\pi^2 \epsilon_{4D}}\exp{\left[ -\frac{1}{2} 
																													\begin{pmatrix} 
																													x & x' & y & y' 
																													\end{pmatrix}
																													\Sigma_{4D}^{-1}
																													\begin{pmatrix} 
																													x\\
																													x'\\
																													y\\
																													y' 
																													\end{pmatrix}
																								\right]}, & \textrm{if} \; x > x_e+Lx' \\
            0 , & \textrm{if} \; x \leq x_e+Lx'
					\end{matrix}
	\right.
\label{eq:Gaussian_density}
\end{equation}
\end{widetext}
where $\Sigma_{4D}^{-1}$ is the inverse of the 4D beam matrix at the detector plane. The resulting transverse profile, $\rho_x(x,y)$, is obtained by integrating over the angular distribution in $x$ and $y$:
\begin{equation}
\label{rhox}
\rho_x(x,y) = \int_{-\infty}^{\frac{x-x_e}{L}} {\rm d}x' \int_{-\infty}^{\infty} {\rm d}y' \rho(x,x',y,y')
\end{equation}
The solution of the integral can be written in the form:
\begin{equation}
\rho_x(x,y) = \frac{1}{2} \rho_0(x,y) \left[1+\textrm{erf}\left( \frac{x-M_x x_e -\tan(\alpha_x) y }{\sqrt{2} \sigma_x} \right)\right],
\label{eq:rhox}
\end{equation}
where $\rho_0(x,y)$ corresponds to the transverse spatial profile of the beam when the knife-edge does not block the beam:
\begin{align}
&\rho_0(x,y) = \frac{Q_0}{2\pi\sqrt{\langle x^2\rangle \langle y^2 \rangle-\langle xy\rangle^2}} \times \nonumber \\
&\exp \left[-\frac{1}{2} 
																													\begin{pmatrix} 
																													x-x_0&y-y_0 
																													\end{pmatrix}
																													\begin{pmatrix} 
																													\langle x^2\rangle &\langle xy\rangle \\
																													\langle xy\rangle & \langle y^2\rangle 
																													\end{pmatrix}^{-1}
																													\begin{pmatrix} 
																													x-x_0\\
																													y-y_0
																													\end{pmatrix}
																												\right]
\label{eq:rho0}
\end{align}

and includes the x-y second moments at the screen that can be retrieved with a fit of the full beam to a 2D Gaussian function. The other three parameters contained in the error function of Eq.~\eqref{eq:rhox}, which depend on the distance between knife-edge and detector plane L,  describe the point-projected image of the knife-edge at the screen: $M_x$ is the magnification factor of the cut edge position; $\sigma_x$ is the cut width or sharpness of the edge at the screen; and $\alpha_x$ is the cut angle of the edge at the screen, which is typically not zero when x-y coupling is present (See Fig~\ref{figure1}(b)). 
Analogous analysis can be made for a knife-edge scan in the vertical direction. In this case, the beam profile at the screen previously intercepted by the knife-edge at the position $y_e$ is:
\begin{equation}
\rho_y(x,y) = \frac{1}{2} \rho_0(x,y) \left[1+\textrm{erf}\left( \frac{y-M_y y_e -\tan(\alpha_y) x }{\sqrt{2} \sigma_y} \right)\right],
\label{eq:rhoy}
\end{equation}

A fit of the beam profile to Eqs.~\eqref{eq:rhox} and \eqref{eq:rhoy} in horizontal and vertical edge scans, respectively, allows for retrieving a total of nine parameters to describe the beam (x-y second moments, magnification, cut width and angle). Adding the beam sizes at the knife edge plane $\{\sigma_{ke,x},\sigma_{ke,y}\}$ obtained from a charge scan using Eq.~\eqref{eq:charge_scan}, the set of data is, in principle, sufficient to reconstruct the ten elements of the 4D transverse beam matrix. For the sake of simplicity, let us write the inverse of the beam matrix as:
\begin{equation}
	\Sigma_{4D}^{-1}=
	\begin{pmatrix}
		a & b & g & h \\
		b & c & l & m \\
		g & l & d & e \\
		h & m & e & f
	\end{pmatrix}
	\label{eq:invS}
\end{equation}
Solving the integral in \ref{rhox} the magnification factor, cut angle and cut width can be espressed in terms of the inverse beam matrix elements as follows:
\begin{align}
M_x &= \frac{1}{1+L \frac{bf-hm}{cf-m^2}} \\
\tan(\alpha_x) &= \frac{em-fl}{bf-hm+\frac{cf-m^2}{L}} \\
\sigma_x &= \frac{\sqrt{f(cf-m^2)}}{bf-hm+\frac{cf-m^2}{L}} \\
M_y &= \frac{1}{1+L \frac{ce-lm}{cf-m^2}} \\
\tan(\alpha_y) &= \frac{bm-ch}{ce-lm+\frac{cf-m^2}{L}} \\
\sigma_y &= \frac{\sqrt{c(cf-m^2)}}{ce-lm+\frac{cf-m^2}{L}} 
\end{align}

The dependence of these parameters from the edge-to-screen distance $L$ becomes very strong when for longitudinal positions of the obstacles around the focal plane. When the beam is focus and knife-edge plane coincide, neither the magnification, the cut angle nor cut width are well-defined, as will be shown in simulations in the following section. 
For this measurement a tightly focused beam is desirable to maximize the magnification at the detector, and to decrease the error in retrieving the local angular spread from fitting the cut width (Eqs.~\eqref{eq:rhox} and \eqref{eq:rhoy}).

Our measurement is actually composed by a series of knife-edge scans at different longitudinal positions around the beam waist. This is particularly important as the data can be combined together in a global fitting strategy to increase the strength of the fit and minimize the systematic/statistics error. The fit is based on the minimization of the squared error between the data $s_i^{(k)}$ and the parametric model $t_i^{(k)}(\Sigma^{\textrm{(recon)}}_{4D},L_0)$, where $\Sigma^{\textrm{(recon)}}_{4D}$ is the 4D transverse beam matrix at the upstream reconstruction point, located at distance $L_0$ from the knife-edge plane. (See Fig.~\ref{figure1}), the subscript $i$ indicates the scan measurement at an arbitrary longitudinal plane $z_i$ and superscript $k$ indicates the measured variable (rms size at edge and detector, magnification factor, cut angle and cut width). The errors are weighted by the uncertainty $\delta s_i^{(k)}$ of each measurement point and they are added into an overall sum:
\begin{equation}
\chi^2\left(\Sigma^{\textrm{(recon)}}_{4D},L_0\right) = \sum_{k,i} \left( \frac{s_i^{(k)}-t_i^{(k)}(\Sigma^{\textrm{(recon)}}_{4D},L_0)}{\delta s_i^{(k)}} \right)^2
\end{equation}

A nonlinear solver routine in MATLAB\cite{fminsearch} is used to find the parameters $(\Sigma^{\textrm{(recon)}}_{4D},L_0)$ that minimize the cost function $\chi^2$. Additional constraints have been included to prevent the solution to be unphysical (i.e. the solution eigenvalues of the beam matrix must be positive). 

Note that to speed up the convergence of the fit it is important to pick a reasonable a starting solution. In our case we estimated beforehand an initial condition for the beam matrix using the reconstruction of the beam profile with the affine mapping (See Appendix). 

\section{\label{sec3}Simulations of the measurement}

\begin{figure*}
\includegraphics[width=.9\textwidth]{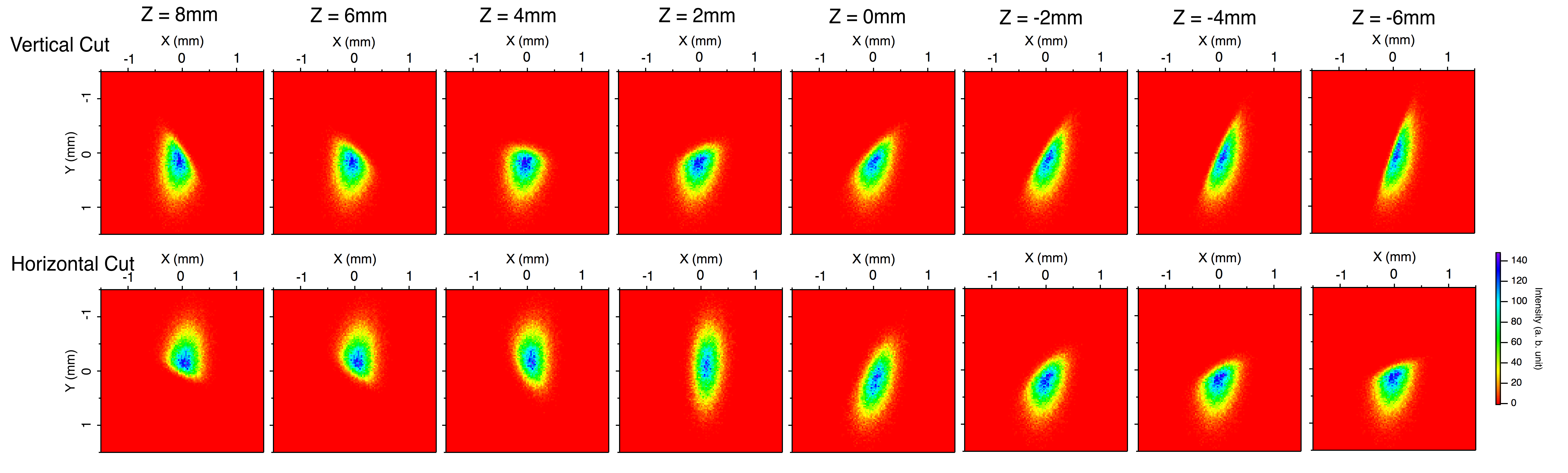}
\includegraphics[width=.9\textwidth]{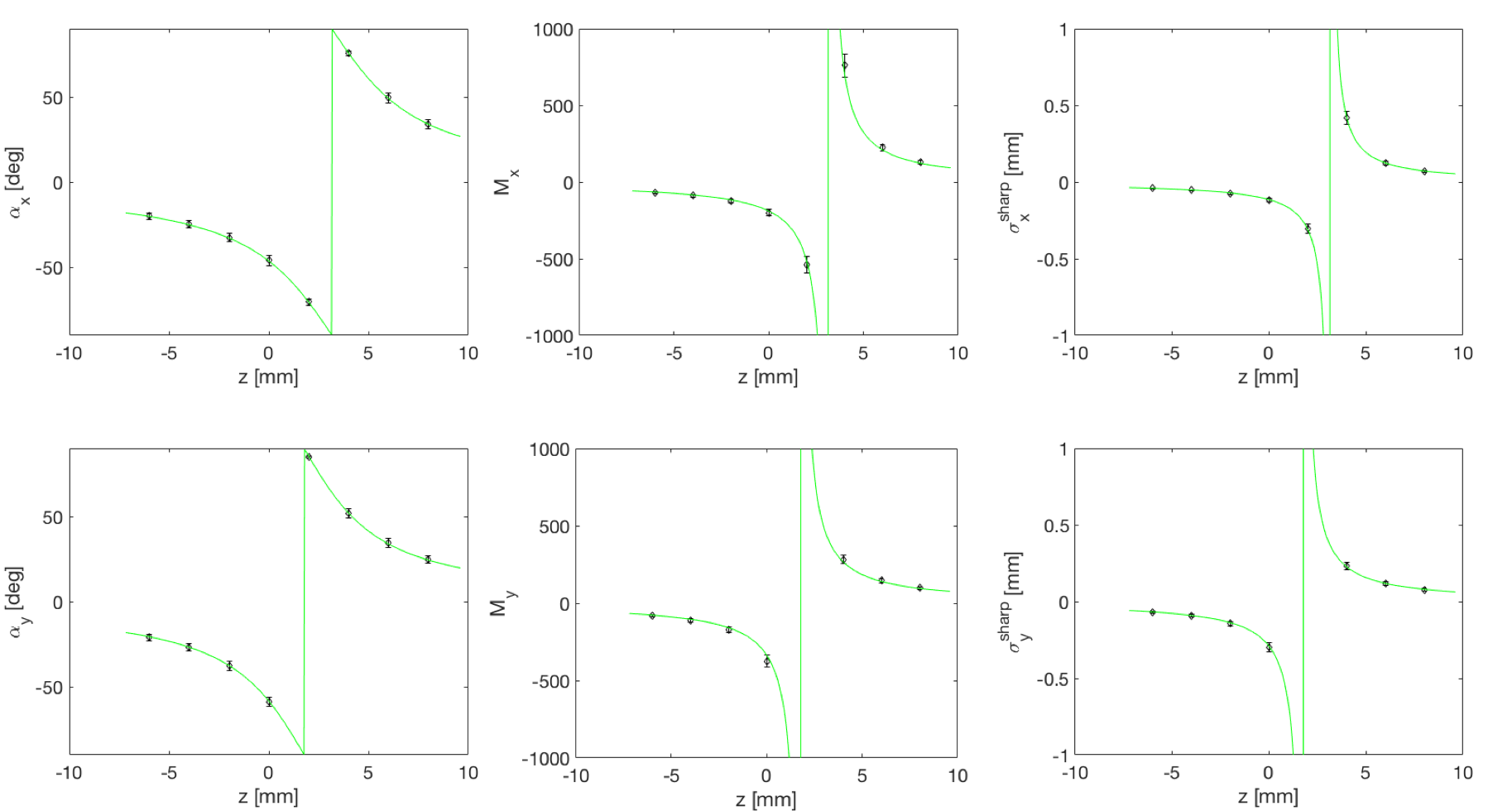}
\includegraphics[width=.8\textwidth]{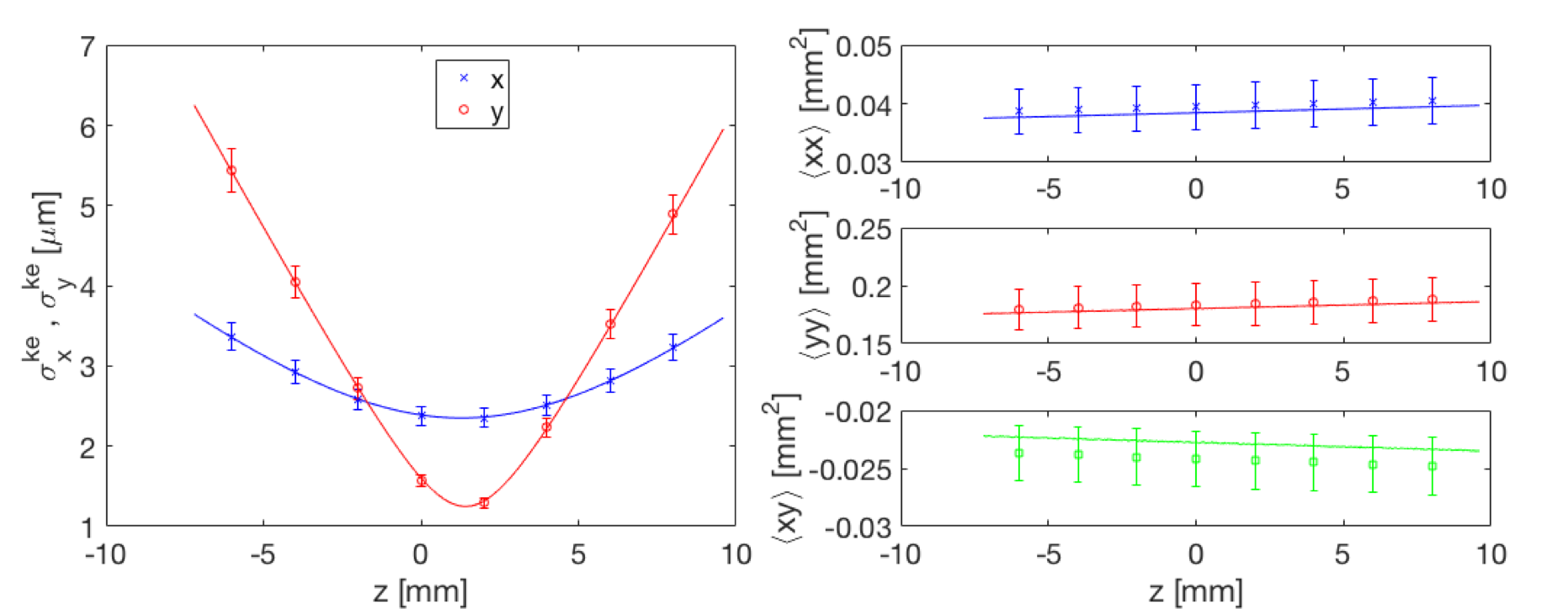}
\caption{\label{figure2} Virtual measurement and its data analysis. (top) Simulated electron beam transverse shape at the detector screen after intercepting a vertical (top row) and horizontal (lower row) semi-infinite plane, for different longitudinal positions of the plane. The \textit{(x,y)} planes are coupled.  Example of curve fitting of virtual data: (middle) cut width, cut angle and magnification; (bottom)  rms beam sizes at the knife-edge and the detector.}
\end{figure*}

\begin{figure*}
\includegraphics[width=.9\textwidth]{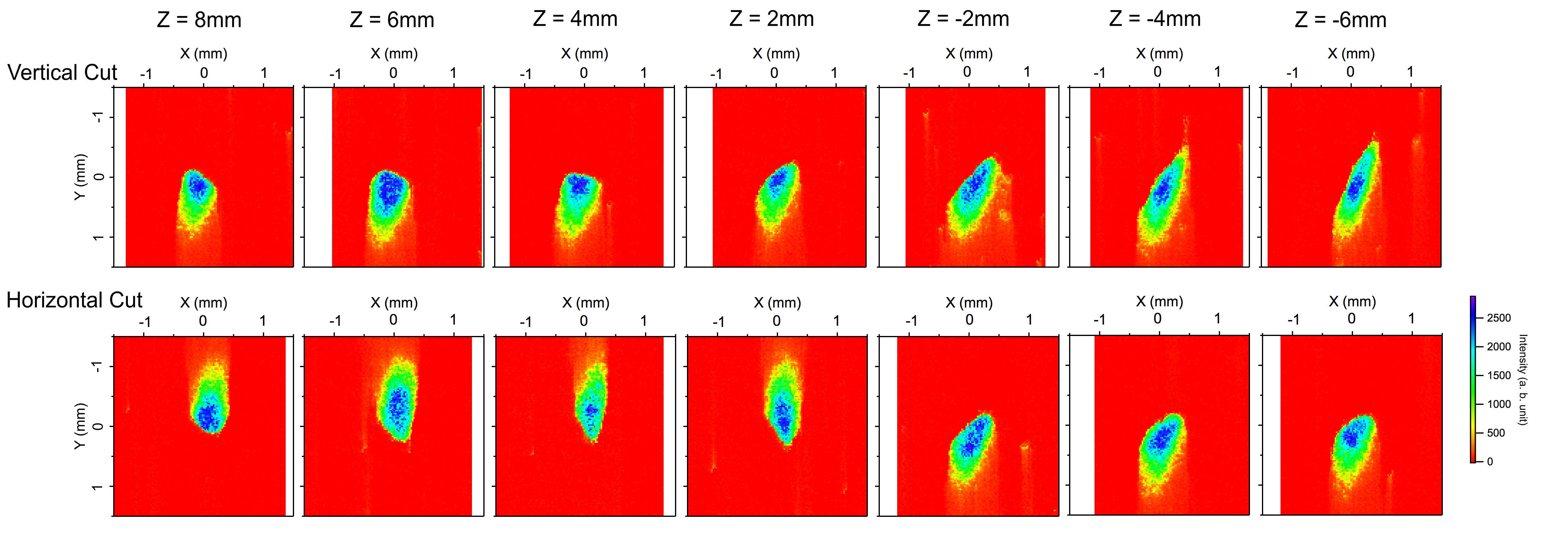}
\caption{\label{figure3} Measured intensity profiles of the half-cut beam over the longitudinal scan, the cut projection rotation is clearly seen}
\end{figure*}

To verify the analysis approach, we performed virtual measurements using a particle tracking code (General Particle Tracer, \cite{GPT}). Figure~\ref{figure2} shows an example of such tests. Starting from real measured data, we derived the initial electron beam matrix parameters and  beam distribution, reported in Table~\ref{table1}.  A random distribution of $10^5$ electrons was generated from the beam matrix,  and then propagated through the knife-edge obstacle and a drift of 0.6 $m$ up to the detector plane (Fig~\ref{figure1}). The knife-edge was simulated using an hard-edge model, i.e. by removing the electrons with $x(y)<0$ for vertical (horizontal) scan configuration. The detector resolution was set to  23.81 $\mu m$/pixel, equal to the camera resolution in the real experimental setup.

Simulations of knife-edge scans are performed at several longitudinal positions along the beam propagation. These span a distance of 14~mm, equal to the total travel distance of our picomotors in experiments. The different positions are labeled from $z = -6 mm$ to $z = 8 mm$ to match the motor encoder readback, with -6~mm being the furthest distance between the lens and the obstacle (about 32~mm). 
To better show the beam behavior in Fig.~\ref{figure2} we report one example of the simulated beam \textit{(x,y)} distributions at the detector after intercepting the obstacle in the center. 
The projection of the edge on the detector undergoes a rotation during the scan for both horizontal and vertical directions; this rotation is due to the presence of x-y correlations in the initial beam matrix.

For each $z$ location the edge was moved along the entire beam with a step size of 250 nm, generating a series of images used in the fitting procedure described in Sect~\ref{sec2} with Eq.~\ref{eq:rhox},\ref{eq:rho0},\ref{eq:rhoy}. The extracted values for cut angle, cut width and magnification, as well as the the rms beam size at the knife-edge plane and detector (Eq.~\ref{eq:charge_scan}) are shown in Fig. \ref{figure2}, error bars are from the fitting confidence interval due to the limited number of particles used. The extracted data points are used as input to the global fitting procedure (see Sect.~\ref{sec2}) to obtain the 4D beam matrix at the reconstruction point. Table I shows a comparison between the original beam matrix and the fitting result, indicating excellent agreement in terms of all matrix elements as well as the projected and 4D transverse emittance. Note that the reconstructed 4D transverse emittance (geometric) is very small ($0.00195  (nm\cdot rad)^2 = (0.044  nm\cdot rad)^2 $) with a relative error of $0.1\%$.

The virtual measurement is an ideal case where the precision is mainly limited by errors due to the limited number of particles used and the sampling resolution at the detector plane. In practical measurements, systematic errors like the calibration error of the camera and error of the distance between knife-edge and the detector need to be considered. Since the knife-edge scan is a form of multi-shot/single optics measurement, beam position jitter and drift at the knife-edge plane over the scan need to be compensated.


\begin{table}%
\caption{Results of the 4D beam matrix and emittance reconstructed from the virtual measurements.}
\label{table1}
\begin{tabular}{c l c c }
\hline
& & Initial & Reconstructed \\
\hline
$\langle x^2 \rangle$ & [$\mu$m$^2$] & $59.83$ &$59.20$\\
$\langle xx' \rangle$ & [$\mu$m mrad] & $-2.432$ &$-2.406$\\
$\langle x'^2 \rangle$ & [mrad$^2$] & $0.1087$ &$0.1079$\\
$\langle y^2 \rangle$ & [$\mu$m$^2$] & $256.5$ &$256.4$\\
$\langle yy' \rangle$ & [$\mu$m mrad] & $-11.38$ &$-11.37$\\
$\langle y'^2 \rangle$ & [mrad$^2$] & $50.76$ &$50.69$\\
$\langle xy \rangle$ & [$\mu$m$^2$] & $11.27$ &$11.18$\\
$\langle xy' \rangle$ & [$\mu$m mrad] & $-0.2045$ &$-0.2085$\\
$\langle x'y \rangle$ & [$\mu$m mrad] & $1.065$ &$1.074$\\
$\langle x'y' \rangle$ & [mrad$^2$] & $-0.06062$ &$-0.06082$\\
\hline
$\epsilon_{4D}(geometric)$ & [(nm rad)$^2$] & $0.001950$ &$0.001948$ \\
$\epsilon_{4D}(normalized)$ & [(nm rad)$^2$] & $0.009642$ &$0.009632$ \\
\hline
\end{tabular}
\end{table}

\section{\label{sec4}application to the HiRES beamline}

In this section we demonstrate the ability of the presented method to reconstruct 4D phase space of a sub-nm\textsuperscript{2} emittance electron beam, generated using the HiRES  beamline at LBNL \cite{HiRES}\cite{HiRES_2}. It employs the APEX \cite{APEX} radio frequency (RF) electron source to provide sub-picosecond electron bunches with repetition rates up to 1 MHz, and an accelerating gradient in excess of 20 MV/m. The electrons are excited via photoemission from a CsK\textsubscript{2}Sb coated copper cathode \cite{KCsSb} by a 150 fs (rms) frequency-doubled Ytterbium-fiber laser. Energy spreads below $ 10^{-4}$ can be achieved depending on charge and pulse length,  with a nominal energy of 735 KeV. 

To deliver an ultra-low emittance beam to the knife-edge scan setup, the HiRES beamline has been optimized as follows. First, the laser spot on the cathode was tightly focused down to 50 um RMS, which minimizes the initial transverse emittance. Second, the combination of the gun solenoid and a following 500 um aperture cuts the high divergence part of the beam and decreases the transverse normalized emittance to 3 nm~\cite{nanoUED}. Then, the beam was transported through the dogleg energy collimator: two sets of quadrupole triplets lenses were utilized to compensate the energy dispersion and minimize the transverse emittance growth due to the energy spread. Finally, the beam went through the 2nd aperture 200 um in diameter about 50~cm upstream of the lens assembly, decreasing the transverse emittance  further into the sub-nm regime.

An in vacuum high-gradient permanent magnet quadrupole (PMQ) lens system~\cite{nanoUED} was utilized to strongly focus the beam. 

For precise reconstruction of the beam, we fabricated microscale knife edge samples with nanometer-level edge sharpness. These consist of a 75 nm Au film thermally evaporated onto a 50 nm SiN membrane suspended over a 0.25 mm x 0.25 mm aperture on a Si support frame. A 10 $\mu$m x 10 $\mu$m square hole was milled through the Au/SiN film with a focused Ga ion beam, providing two vertical and two horizontal knife edges. The peak-to-peak edge roughness was measured with SEM to be less than 10 nm. The polycrystalline Au layer is thick enough to filter out the overlapping portion of the beam by scattering. No transmission through the Au was observed within the detection sensitivity of the instrument. This provides sharp, high-contrast beam cuts as demonstrated in Fig.~\ref{figure3}. 

A scintillator screen (Ce:YAG, 25 mm diameter and 100~$\mu m$ thick) was used to image the beam at 0.6~m downstream from the knife-edge. The scintillator was then imaged onto the CCD of an intensified camera (Princeton Instruments PI-MAX4), the detector assembly calibration was determined to be 23.81 um/pixel.

In order to apply the model described in Sect.\ref{sec2}, we performed knife-edge measurements at multiple longitudinal positions along the beamline. 
While the knife-edge obstacle was moved transversely over the beam, the longitudinal scan was accomplished by translating the entire PMQ lens assembly rather then the obstacle. With reference to Fig.~\ref{figure1}, the distance $L$ between the knife-edge and the detector was kept fixed, while the distance $L0+z$ varied. In principle such a procedure, forced by our particular experimental setup, does not maintain a constant input beam matrix $\Sigma_{4D}$ during the measurements. 
Nevertheless for a collimated beam, the beam matrix before the lens is very slowly changing and can be with a good approximation be assumed constant.



Figure \ref{figure3} shows the experimental data along a total longitudinal scan range of 15~mm. As in the case of the virtual measurements presented above,we only report one image for each longitudinal position. The cut angle rotates as function of edge longitudinal position, as predicted by the theoretical model.

\begin{figure}
\includegraphics[width=8cm]{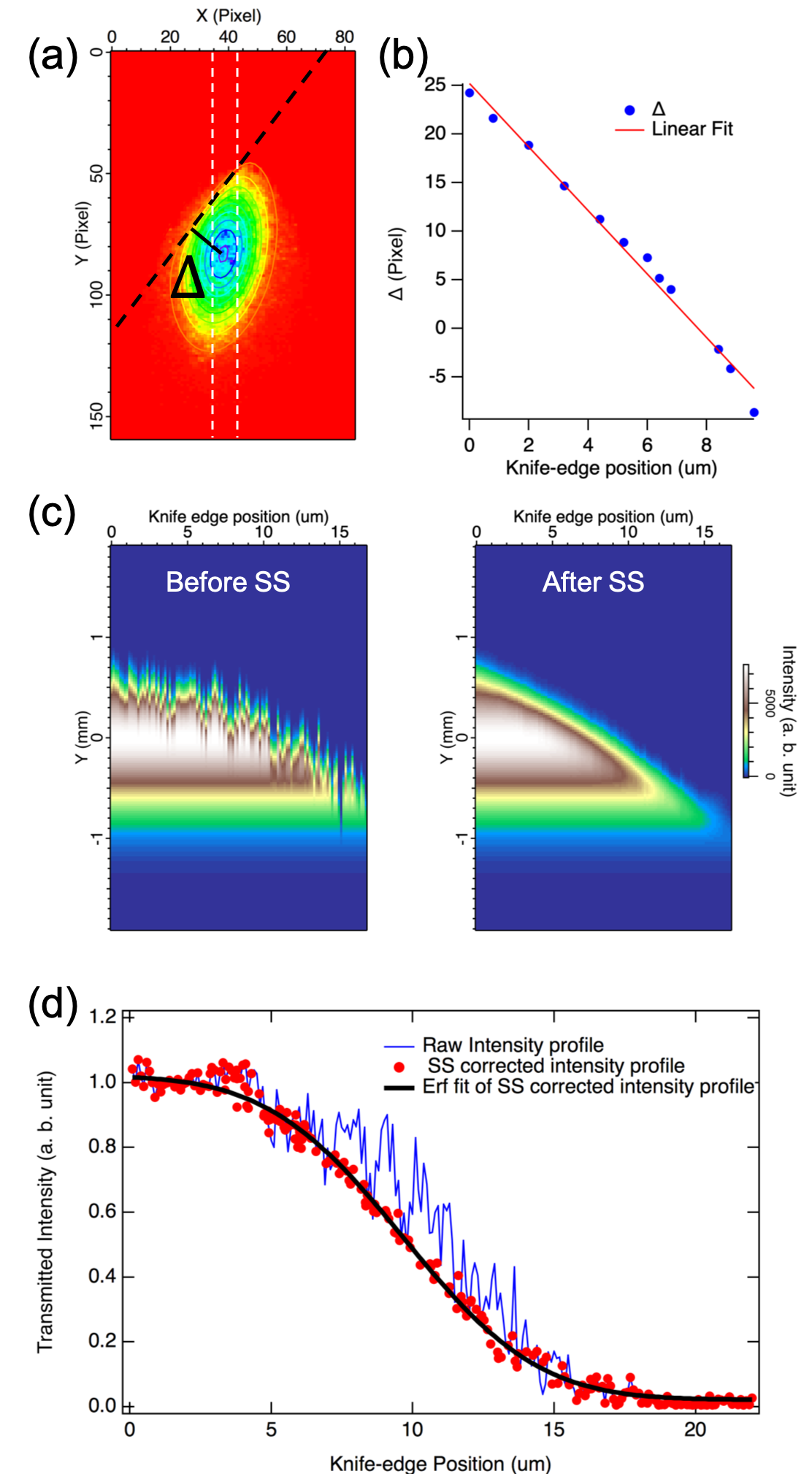}
\caption{\label{figure4} Space-stamping to correct spatial jitter (a) typical knifed beam on detector, $\Delta$ is defined as the distance between cut and Gaussian centroid. (b) Linear fit of $\Delta$ using transmitted profiles with similar Gaussian centroids. (c) Intensity profile in the white dashed box region in (a) over the knife-edge scan without (left) and with (right) SS correction. (d) Comparison of the transmitted intensity profile with and without SS correction.}
\end{figure}

\subsection{\label{sec5_1}Spatial jitters compensation}

Stroboscopic characterization and application of ultrafast, nm-scale electron beams requires exquisite beam transverse and longitudinal stability. Parameter fluctuations in the experimental setup, such as power supply current instability, ground vibrations, amplitude and phase fluctuations in the radiofrequency levels can negatively impact the short- and long-term electron beam jitter.
To circumvent the loss of temporal resolution due to the time-of-arrival pump-probe jitter, time stamping techniques have been developed and widely used over the years in FELs and UED setups \cite{Harmand}\cite{Gao}\cite{zhao_terahertz_2018}.
In the following we apply a similar concept to the spatial coordinate to compensate for beam transverse instabilities. We call the technique space stamping (SS), in analogy with the temporal coordinate. 




The SS approach is based on utilizing the beam centroid information from each image to correct for the spatial beam fluctuations during the knife-edge scan. The correction is applied as follows: first, for each scan step at a fixed longitudinal position z, the transmitted beam profile was fitted with Eq. \ref{eq:rhox} and \ref{eq:rhoy}. Second, the distances $\Delta$ between the Gaussian centroid (the beam position) and knife-edge cut were retrieved (Fig.~\ref{figure4}(a)). Under the assumption of linear optics, the distance $\Delta$ is directly related to the knife-edge position $\Delta$x on the beam through a scaling factor $t$. Such scale is determined by first ordering all the images in terms of Gaussian centroid, then selecting for each knife edge position the images with centroid differing by less than 1 pixel from the median of the distribution, and finally performing a linear fit to find $t$ on the selected profiles (Fig.~\ref{figure4}(b)). Lastly, the knife-edge positions $\Delta$x of every image for all the scan steps are re-assigned according to the cut-centroid distance $\Delta$ and the scale $t$.
Fig.~\ref{figure4}(c) and (d) show the dataset for a particular knife-edge scan uncorrected and SS-corrected, showing clear suppression of noise and higher goodness of fit after correction.
Such technique therefore can be efficiently applied to remove spatial fluctuations. On the other hand its bandwidth is limited to pointing jitters slower than the acquisition frame rate (1-few Hz).

\subsection{\label{sec5_3}Results and Discussion}

A fit of the complete set of data retrieved from the beam profiles of the knife-edge scans allows for the reconstruction of the 4D beam matrix following the theoretical model detailed in Section~\ref{sec2}. 

The 4D beam matrix reconstruction is made on the lens output plane (see Fig.~\ref{figure1}). This beam matrix is related to the one on the screen, $\Sigma^{\textrm{(screen)}}_{4D}$, by a drift transport matrix of distance $L+L_0+z$, where $z$ is the position of the lens, measured by the translation stage it is mounted on, and $L_0$ is the distance between the lens and the knife-edge when the position is set at $z=0$. Because $L_0$ could not be measured with good accuracy, we included it as parameter in the overall fitting reconstruction.

\begin{figure*}
\includegraphics[width=.9\textwidth]{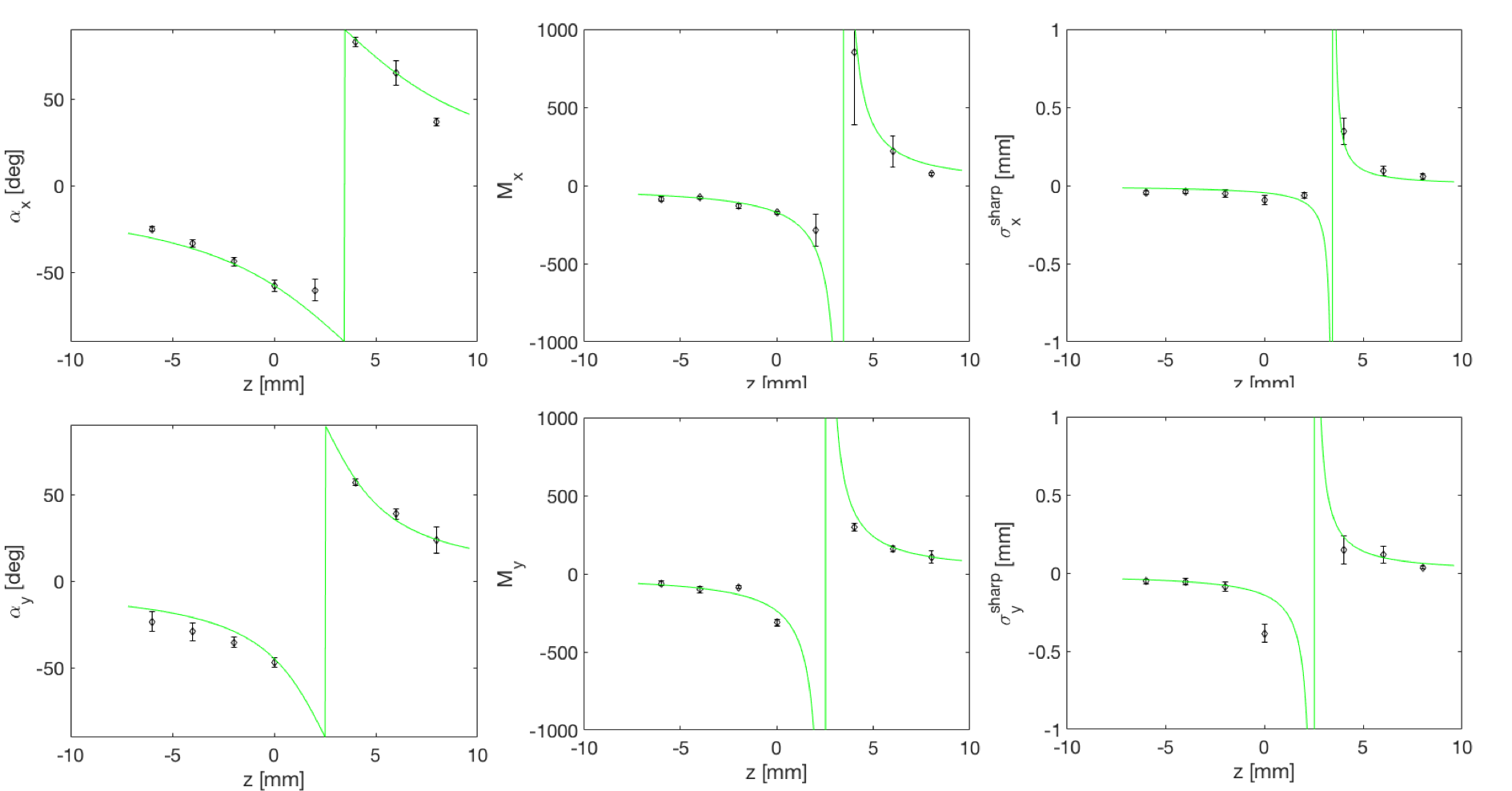}
\includegraphics[width=.8\textwidth]{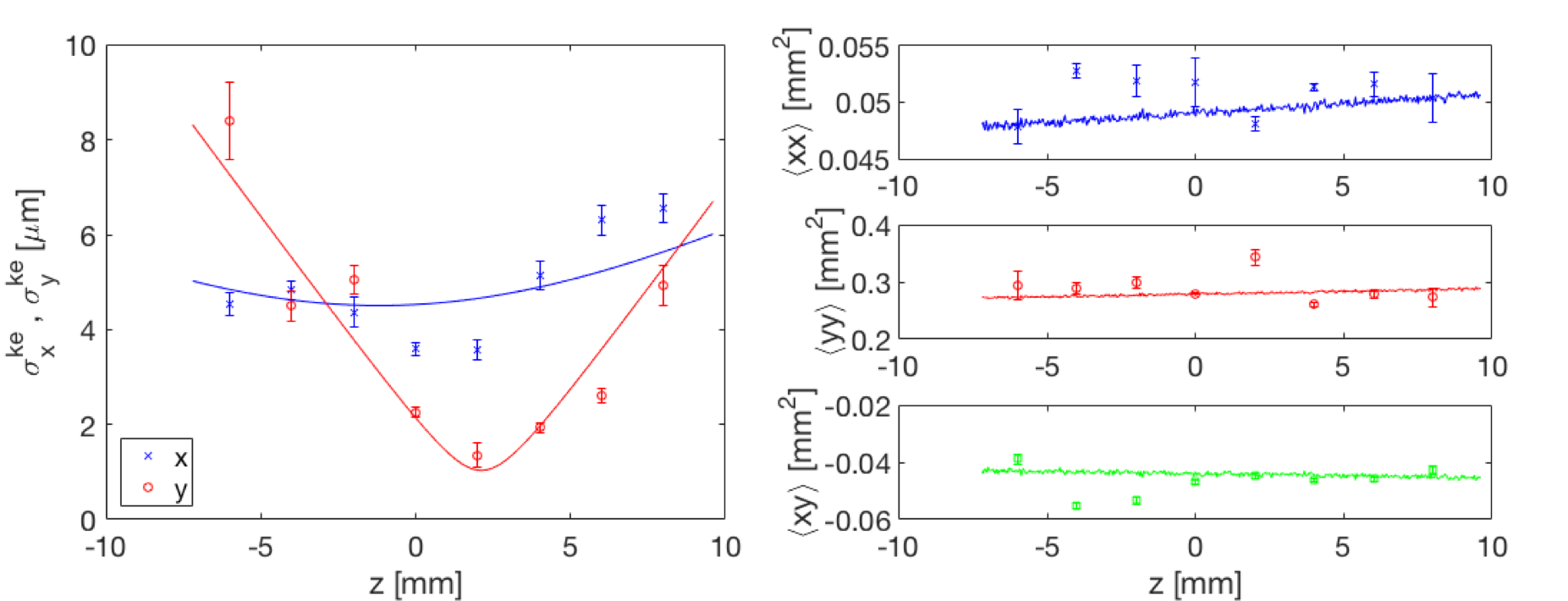}
\caption{\label{figure6} Results of the global fitting to the measured edge data (top) and the rms sizes at the knife-edge and the detector planes (bottom).}
\end{figure*}

The solution of the beam matrix reconstruction is summarized in Table~\ref{table2}, while the result of the global fit is shown in Fig.~\ref{figure6}. 

\begin{table}%
\caption{Results of the 4D beam matrix reconstruction and emittance. Errors are corresponding to standard deviation given by the Monte Carlo simulation}
\label{table2}
\begin{tabular}{c l c}
\hline
$\langle x^2 \rangle$ & [$\mu$m$^2$] & $73.4\pm0.3$ \\
$\langle xx' \rangle$ & [$\mu$m mrad] & $-2.69\pm0.01$ \\
$\langle x'^2 \rangle$ & [mrad$^2$] & $0.1365\pm0.0004$ \\
$\langle y^2 \rangle$ & [$\mu$m$^2$] & $420\pm0.9$ \\
$\langle yy' \rangle$ & [$\mu$m mrad] & $-18.14\pm0.02$ \\
$\langle y'^2 \rangle$ & [mrad$^2$] & $0.786\pm0.002$ \\
$\langle xy \rangle$ & [$\mu$m$^2$] & $40.30\pm0.08$ \\
$\langle xy' \rangle$ & [$\mu$m mrad] & $-1.42\pm0.01$ \\
$\langle x'y \rangle$ & [$\mu$m mrad] & $2.38\pm0.01$ \\
$\langle x'y' \rangle$ & [mrad$^2$] & $-0.1161\pm0.0003$ \\
\hline
$\epsilon_{4D}(geometric)$ & [(nm rad)$^2$] & $0.0029\pm0.0013$ \\
$\epsilon_{4D}(normalized)$ & [(nm rad)$^2$] & $0.0144\pm0.0065$ \\
\hline
\end{tabular}
\end{table}

The precision of the emittance measurement will be determined by the resulting uncertainties of the global fitting parameters. In order to assess the performance of the knife-edge scan technique, we carried out a sensitivity study of the fitting routine to the statistical errors observed during the measurement of beam-edge profiles.

We performed Monte Carlo simulation in which 5000 datasets are randomly generated based on the profile measurements and their uncertainties, following a Gaussian distribution centered at $s_i^{(k)}$ and with a standard deviation of $\delta s_i^{(k)}$. Every dataset is processed as an input of the global fitting routine. A statistical analysis of the resulting reconstructions allows for an estimation of the standard deviation of the measured emittance. 

\begin{figure*}
\includegraphics[width=0.8\textwidth]{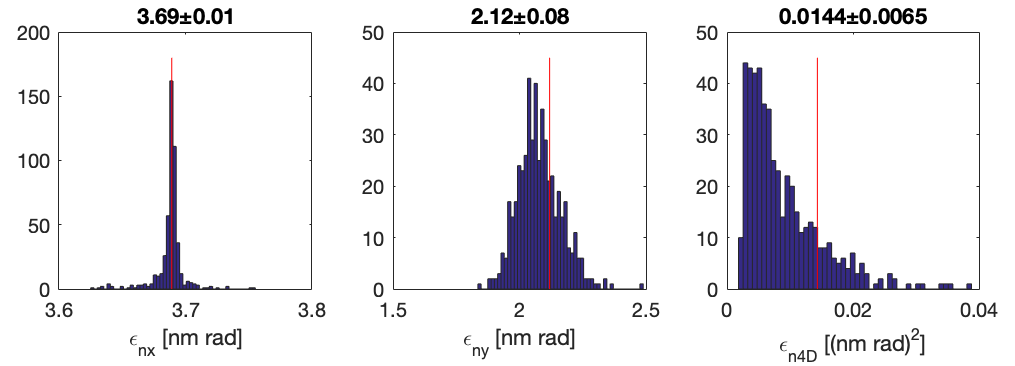}
\caption{\label{figure7} Statistical distribution of the normalized apparent and 4D emittance values in the Monte Carlo simulation}
\end{figure*}

Table~\ref{table2} includes the standard deviation of the reconstructed beam matrix elements and emittances. Figure~\ref{figure7} shows the statistical distribution of the normalized emittance values obtained from the Monte Carlo simulation. The emittance along the horizontal direction $\epsilon_{nx}$ was calculated from the reconstructed matrix using the following equation: 
\begin{equation}
\epsilon_{nx} = \beta\gamma\sqrt{\langle x^2\rangle\langle x'^2\rangle-\langle xx'\rangle^2}
\label{eq:epsilonx}
\end{equation}
and similar for $\epsilon_{ny}$. The 4D emittance was calculated from Eq.~\eqref{4Demittance}. The large difference (about 2 orders of magnitude) between the product $\epsilon_{nx}\epsilon_{ny}$ and $\epsilon_{n4D}$ confirms the presence of strong correlations between the two transverse planes, and demonstrates the importance of a full four-dimensional reconstruction. Indeed our experimental results using such focused beams~\cite{nanoUED} for ultrafast electron diffraction show spatial resolution performance in line with the higher phase space density and transverse coherence length obtained from the four-dimensional emittance value.  




\begin{figure*}
\includegraphics[width=1.0\textwidth]{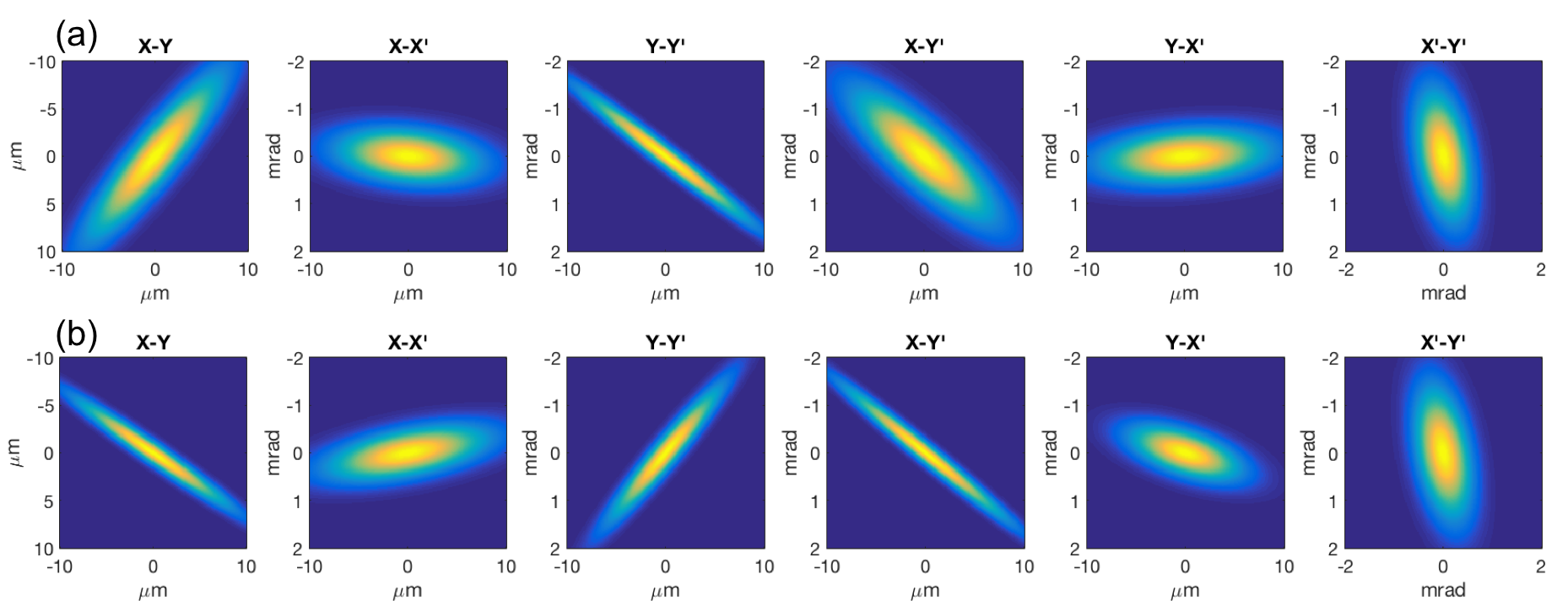}
\caption{\label{figure8} Reconstructed phase space projections at z = -4 mm (a) and z = 6 mm (b)}
\end{figure*}

Such technique, similarly to the more standard quadrupole-scan techqnique, can be used in the event that space charge effects can be neglected and a simple drift transport assumed for the electrons between the knife-edge obstacle and the detector plane. Here we estimate the pulse charge expected from tightly focused electron beams with picometer-scale emittance. The 6D beam brightness is defined as:
\begin{equation}
B_{6D}=\frac{eN_{tot}}{(\sqrt{2\pi})^3\epsilon^2_n\sigma_t(\sigma_E/mc^2)}
\label{eq:B6d}
\end{equation}
taking into account the peak current, transverse normalized emittance as well as the energy spread, and the beam is assumed to have no correlated time-energy chirp. . By using state-of-art radiofrequency-based electron sources, 6D beam brightness values as high as $10^{17} A/m^2$ can be achieved\cite{SourceBrightness}. For an electron beam with $0.0144 (nm rad)^2$ 4D normalized transverse emittance, 300 fs FWHM pulse length and $10^{-3}$ energy spread, the calculated average pulse charge is $<10$ electrons/pulse. Thus, for small enough emittance values and state-of-art photocathode brightness values, the space charge effect can be neglected and the reported analysis is applicable.


Although the space-stamping technique remove a large part of the transverse beam fluctuations, spatial fluctuations due to beamline machanical vibrations at frequencies  above 10 Hz still remain. This is the major cause of the uncertainty in our measurements and it constitutes in general the main limitation to measuring small beam sizes and small emittance values. Another main component of the measurement error is represented by the coupling between different planes. For strongly coupled 4D phase spaces, the precision in measuring the coupling and the projections along the different coordinates becomes exponentially important to obtain a reliable value for the 4D volume \cite{Jorge}. The x-y correlation can result from a non symmetric initial laser profile, the x-y coupling from gun solenoid lens, permanent magnet-based lens rolling/tilting error, etc. 
Nevertheless the presented results show the potential of the knife-edge scan technique to measure ultra-low transverse emittances. 
Fig.~\ref{figure8} shows the reconstructed phase space projections around the beam waist using the obtained beam matrix, the strong coupling between different planes was unveiled. The x-y' and y-x' correlations were the main cause of the cut rotation behavior observed at the detector screen. When the knife edge was not inserted, full beam shape at the detector was dominated by the x'-y' distribution.

\section{\label{sec6}Conclusion and outlook}

The development of ultrahigh brightness, low emittance electron sources calls for highly precise beam measurement techniques. In this work we have presented a novel technique for measurement of the 4D beam matrix of tightly focused electron beams with ultralow emittance values. Previous measurements reported emittances in the range of few nanometers, or $\epsilon_{4D}\geq10~nm^2$. The proposed technique has been successfully tested to characterize complex evolution around the waist of beams with 4D normalized emittances 100 times smaller, down to $0.0144 (nm\cdot rad)^2$.

A detailed theoretical model has been described and validated via virtual measurement performed by making use of the GPT  particle tracking code \cite{GPT}.We then apply the knife-edge scan technique at the measurement of sub-micrometer beams generated at the HiRES beamline at LBNL. A detailed data processing procedure was developed, including space stamping and affine mapping for guessing the initial parameter values in the global fitting.

The knife-edge routine presented here has a few main advantages over other techniques. First, the scans can be made with arbitrarily small step size, only limited by the precision of the particular piezomotor used in the measurements (in the few $nm$ range). Also, pointing jitters can be compensated by space stamping if the frame rate during acquisition is high enough; this is especially effective in high repetition rate setups with high electron flux. Second, the image analysis and global fitting procedure allow to reconstruct the behaviour of the full transverse phase space. Such information is crucial in the optimization of tightly focused electron beam and strong lens systems.


Ultrafast electron beams with emittance values below the nanometer and beam size in the sub-micrometer regime are finding wider applicability, from ultrafast electron diffraction and microscopy, to Dielectric Laser Acceleration~\cite{England} and external electron injection in laser-plasma accelerators. 
High brightness electron source R\&D projects are being pursued in order to get higher brightness from the emission surface \cite{BE}\cite{KCsSb} and increase the accelerating field, output energy together with the average electron flux \cite{APEXII}. We believe that techniques like the one presented here will become crucial as the R\&D on the sources progresses.


\begin{acknowledgments}
We thank D. Arbelaez and T. Luo for help on magnetic measurements of the PMQ lens. We acknowledge L. Doolittle and G. Huang for support on LLRF system. The knife-edge scan hardware was acquired thanks to funds provided by LBNL through the Laboratory Directed Research and Development plan (LDRD). This work was supported by the U.S. Department of Energy under Contract No. DE-AC02- 05CH11231. This LBNL-UCB-UCLA collaboration was supported by STROBE: A National Science Foundation Science and Technology Center under Grant No. DMR 1548924, which provided funding for DD. JGN acknowledges support from the Center for Bright Beams, NSF award PHY 1549132.
\end{acknowledgments}

\appendix*

\section{\label{affine}affine transformation to obtain the initial guess of beam matrix elements}
\begin{figure}
\includegraphics[width=8cm]{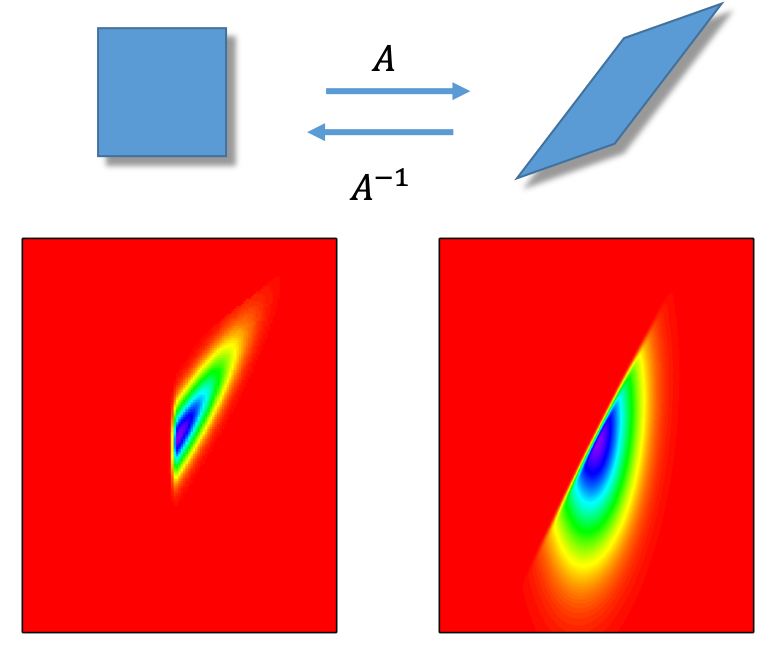}
\caption{\label{figure5} Affine transformation (top), simulated intensity profile of a half-cut beam at the knife-edge plane (bottom left) and intensity profile on the detector (bottom right), the transformation matrix $A$ is given by Eq.~\ref{eq:affine}}
\end{figure}

The initial guess for the fitting parameters is a key step in the data analysis process, as the final convergence of the algorithm is strongly dependent upon the particular choice made.
To obtain a proper initial guess (Section \ref{sec5_3}), we developed a procedure to reconstruct the beam profile at the knife-edge plane based on affine transformation. The beam profile in the \textit{(x,y)} plane at the knife edge can be connected to the profile at the detector by a $2 \times 2$ matrix $A$  including shear, rotation and scaling (see Fig. \ref{figure5}):

\begin{equation}
\rho_{detector}(x,y) = \rho_{knife-edge}(A^{-1}(x,y)) 
\label{eq:affinemap}
\end{equation}

The projection image of the horizontal and vertical knife-edge cut is a direct sampling of the affine map: 

\begin{equation}
	A=
	\begin{pmatrix}
		\frac{M_x}{1-\tan(\alpha_x)\tan(\alpha_y)} & \frac{M_y\tan(\alpha_x)}{1-\tan(\alpha_x)\tan(\alpha_y)}  \\
		\frac{M_x\tan(\alpha_y)}{1-\tan(\alpha_x)\tan(\alpha_y)} & \frac{M_y}{1-\tan(\alpha_x)\tan(\alpha_y)}  
	\end{pmatrix}
	\label{eq:affine}
\end{equation}

Where $M_{x(y)}$,$\alpha_{x(y)}$ are respectively the magnification and projected cut angle. The beam profile at the knife-edge plane is given by:

\begin{equation}
\rho_{knife-edge}(x,y) = \rho_{detector}(A(x,y)) 
\label{eq:affinemaprev}
\end{equation}


Fig.\ref{figure5} shows a simulated half-cut beam profile on the detector plane and the reconstructed beam profile at the knife-edge plane using the affine transformation. Note that, 1) affine matrix defines a transformation in \textit{(x,y)} plane (not \textit{(x,x')}), and 2) the profile retrieved from Eqs.~\eqref{eq:affinemaprev} is a backtracking result which does not take into account the 4D beam emittance, and it is therefore less and less accurate approaching the beam waist. 

By applying affine mapping one can obtain the beam profiles at different longitudinal positions. The resulting matrix elements $\langle xx \rangle$, $\langle yy \rangle$ and $\langle xy \rangle$ over the scan can be used to solve the drift transfer problem \cite{Prat}. Given an estimated measurement of $L_0$, the fit to each rms sizes retrieves all beam matrix elements at the reconstruction point except for $\langle xy' \rangle$ and $\langle x'y \rangle$, which only the sum is obtained. Choosing these values appropriately as the starting set of parameters, based on how they fit to the actual data, allows for the global fitting routine to converge. 



\end{document}